\documentclass[twocolumn,showpacs,preprintnumbers,amsmath,amssymb,floatfix,aps]{revtex4}
\usepackage{graphicx}
\usepackage{dcolumn}
\usepackage{bm}

\begin{document}
\title{Comparative Raman Studies of Sr$_2$RuO$_4$, Sr$_3$Ru$_2$O$_7$ and
Sr$_4$Ru$_3$O$_{10}$}
\author{M. N. Iliev$^1$,  V.~N.~Popov$^2$, A.~P.~Litvinchuk$^1$,
M.~V.~Abrashev$^2$, J.~B\"{a}ckstr\"{o}m$^3$, Y.~Y.~Sun$^1$,
R.~L.~Meng$^1$, and C. W. Chu$^{1,4}$} \affiliation{$^1$Texas
Center for Superconductivity and Advanced
   Materials, and Department of Physics, University of Houston,
   Houston, Texas 77204-5002\\
$^2$Faculty of Physics, University of Sofia, 1164 Sofia,
Bulgaria\\
$^3$Department of Applied Physics, Chalmers University of
Technology, S-41296 G\"{o}teborg, Sweden\\
$^4$ Hong Kong University of Science and Technology, Hong Kong,
China}
\date{\today}
\begin{abstract}
The polarized Raman spectra of layered ruthenates of the
Sr$_{n+1}$Ru$_n$O$_{3n+1}$ ($n=1,2,3$) Ruddlesden-Popper series
were measured between 10 and 300~K. The phonon spectra of
Sr$_3$Ru$_2$O$_7$ and Sr$_4$Ru$_3$O$_{10}$ confirmed earlier
reports for correlated rotations of neighboring RuO$_6$ octahedra
within double or triple perovskite blocks. The observed Raman
lines of $A_g$ or $B_{1g}$ symmetry were assigned to particular
atomic vibrations by considering the Raman modes in simplified
structures with only one double or triple RuO$_6$ layer per unit
cell and by comparison to the predictions of lattice dynamical
calculations for the real $Pban$ and $Pbam$ structures. Along with
discrete phonon lines, a continuum scattering, presumably of
electronic origin, is present in the $zz$, $xx$ and $xy$, but not
in the $x'y'$ and $zx$ spectra. Its interference with phonons
results in Fano shape for some of the lines in the $xx$ and $xy$
spectra. The temperature dependencies of phonon parameters of
Sr$_3$Ru$_2$O$_7$ exhibit no anomaly between 10 and 300~K where no
magnetic transition occur. In contrast, two $B_{1g}$ lines in the
spectra of Sr$_4$Ru$_3$O$_{10}$, corresponding to oxygen
vibrations modulating the Ru-O-Ru bond angle, show noticeable
hardening with ferromagnetic ordering at 105~K, thus indicating
strong spin-phonon interaction.
\end{abstract} 
\maketitle
\section{Introduction}
The properties of layered ruthenates Sr$_{n+1}$Ru$_n$O$_{3n+1}$
$(n=1,2,3$), known as the Ruddlesden-Popper  series, exhibit
strong dependence on the number of RuO$_6$ octahedral layers.
Sr$_2$RuO$_4$ ($n=1$) is $p$-wave
superconductor,\cite{maeno1,braden1} Sr$_3$Ru$_2$O$_7$ ($n=2$) is
nearly ferromagnetic (enhanced paramagnetic) metal,\cite{ikeda1}
whereas Sr$_4$Ru$_3$O$_{10}$ ($n=3,\ T_C=105$~K) is a
ferromagnetic metal.\cite{crawford1,cao1} There are indications
that the variations with $n$ of the magnetic and transport
properties of Sr$_{n+1}$Ru$_n$O$_{3n+1}$ are partly related to the
structural distortions in Sr$_3$Ru$_2$O$_7$ and
Sr$_4$Ru$_3$O$_{10}$. While the structure of Sr$_2$RuO$_4$ is
tetragonal ($I4/mmm$, No.139, Fig.1) and the Ru-O-Ru angle in the
$ab$ plane is 180$^\circ$, the structures of Sr$_3$Ru$_2$O$_7$
($Pban$, No.50) and Sr$_4$Ru$_3$O$_{10}$ ($Pbam$, No. 55) are
orthorhombic due to correlated rotations about the $c$-axis of the
neighboring corner-sharing octahedra within each layer of the
double or triple perovskite
blocks.\cite{ikeda1,huang1,crawford1,cao1} These rotations result
in decrease of the Ru-O-Ru angle in the $ab$ plane to 166$^\circ$
for Sr$_3$Ru$_2$O$_7$ and 169$^\circ$ for the outer layers and
158$^\circ$ for the middle layers of Sr$_4$Ru$_3$O$_{10}$,
respectively. The Sr-based ruthenates, however, are less distorted
than corresponding Ca-based compounds. Indeed, besides being
rotated around the $c$ axis, the RuO$_6$ octahedra in
Ca$_2$RuO$_4$ ($Pbca$, No.61)\cite{braden1} and Ca$_3$Ru$_2$O$_7$
($A2_1ma$, No.36) \cite{cao2} are also tilted around an axis lying
in the RuO$_2$ plane.

The coupling among the charge, lattice and spin degrees of freedom
in (Ca,Sr)$_{n+1}$Ru$_n$O$_{3n+1}$ compounds has been subject of
several magnetotransport\cite{ikeda1,cao1},
pressure\cite{shaked1,snow1} and
Raman\cite{snow1,liu1,sakita1,rho1} studies. The polarization-,
temperature-, pressure- and substitution-dependent Raman spectra
allowed observation of the two-magnon scattering, opening of the
spin gap, and pressure- and substitution-induced variations in
metal-insulator transition in Ca$_2$RuO$_4$\cite{snow1},
Ca$_{2-x}$Sr$_x$RuO$_4$\cite{rho1} and
Ca$_3$Ru$_2$O$_7$\cite{liu1,snow1}, as well as spin gap and strong
direction-dependent electron-phonon interaction in
Sr$_2$RuO$_4$\cite{sakita1}. The latter studies illustrated the
ability of Raman scattering to provide information on the
interplay of spin, charge, and lattice degrees of freedom. To our
knowledge, there are yet no reports on the Raman spectroscopy of
double-layer Sr$_3$Ru$_2$O$_7$ and triple-layer
Sr$_4$Ru$_3$O$_{10}$. The Raman spectra of these materials and
their variations with temperature are of definite interest as they
contain information about the local structure, electron-phonon,
spin-phonon interactions and their variations with the number of
RuO$_6$ layers. An essential precondition for correct analysis and
understanding of complex structure-properties relationships is the
assignment of the Raman lines to particular atomic motions.

\begin{figure*}
\includegraphics[width=5in,angle=0]{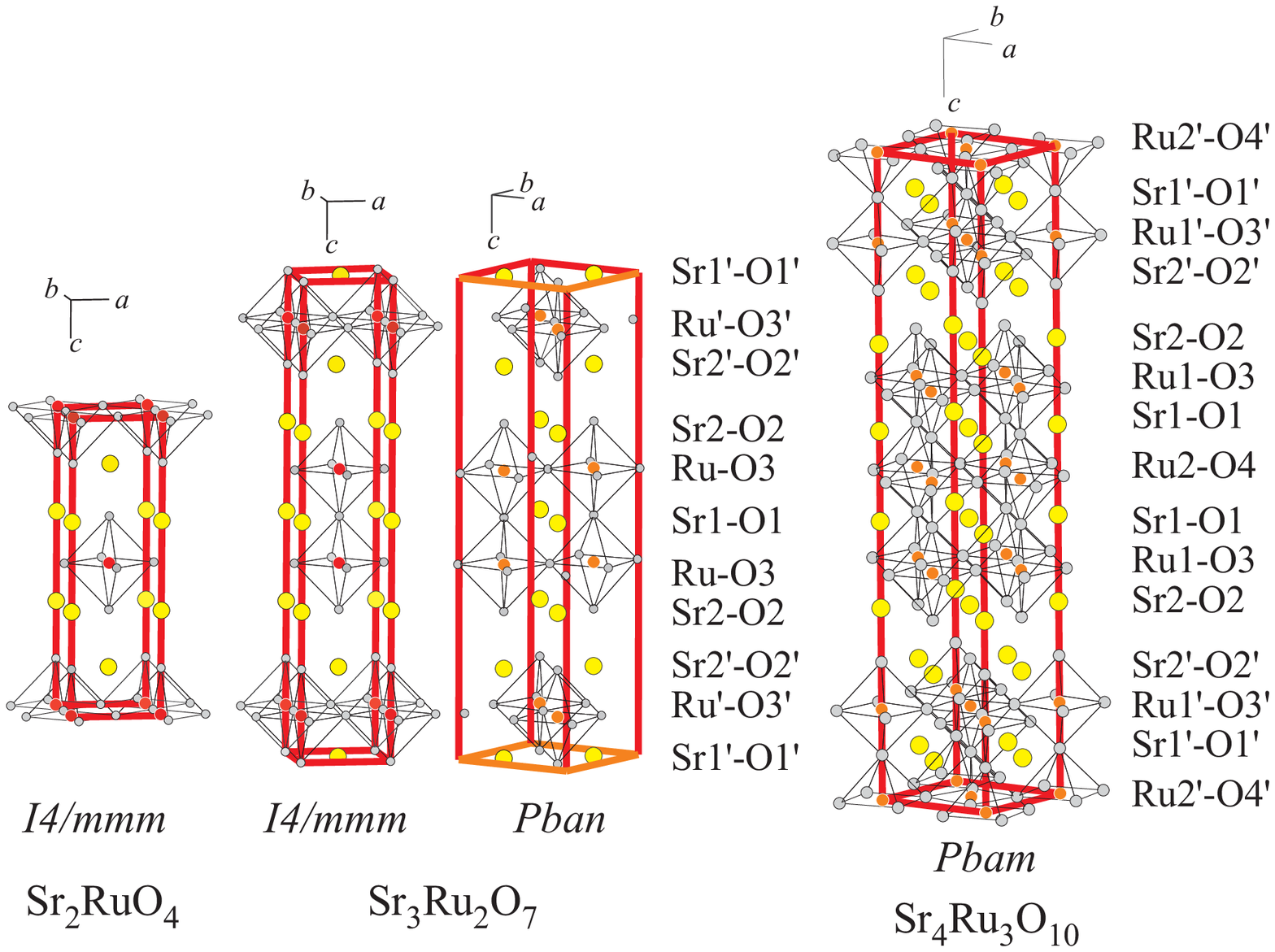}
\caption{Units cells of Sr$_2$RuO$_4$ ($I4/mmm$),
Sr$_3$Ru$_2$O$_7$ ($I4/mmm$ and $Pban$) and Sr$_4$Ru$_3$O$_{10}$
($Pbam$).}
\end{figure*}

In this work we present results of comparative polarization- and
temperature-dependent Raman studies of Sr$_2$RuO$_4$,
Sr$_3$Ru$_2$O$_7$ and Sr$_4$Ru$_3$O$_{10}$ between 10 and 300~K.
The spectra of Sr$_3$Ru$_2$O$_7$ and Sr$_4$Ru$_3$O$_{10}$ provide
clear evidence for a structure containing correlated rotations of
RuO$_6$ octahedra within perovskite blocks. On the basis of their
symmetry, considering corresponding modes in simplified tetragonal
structures, containing RuO$_6$ rotations, and by comparison to the
predictions of lattice dynamical calculations (LDC) for the real
orthorhombic $Pban$ and $Pbam$ structures, the observed Raman
lines of Sr$_3$Ru$_2$O$_7$ and Sr$_4$Ru$_3$O$_{10}$ are assigned
to definite phonon modes. Except for phonon lines, a continuum
scattering, presumably of electronic origin, with components of
$A_g$ ($xx$ and $zz$) and $B_{1g}$ ($xy$) symmetry is present in
the whole temperature range. Its interference with phonons results
in Fano shape for some of the lines in the $xx$ and $xy$ spectra.
Phonon anomalies related to magnetic ordering at T$_C=105$~K are
observed in the temperature-dependent spectra of
Sr$_4$Ru$_3$O$_{10}$.

\section{Samples, Experimental and Lattice Dynamical Calculations}

Rectangular platelet-like single crystals of
Sr$_{n+1}$Ru$_n$O$_{3n+1}$ ($n=1,2,3$) with typical size
1$\times$1$\times$0.2~mm$^3$ were grown using  SrCO$_3$ and
RuO$_2$ as starting materials and SrCl$_2$ as a flux. The
compositions Sr$_2$RuO$_4$, Sr$_3$Ru$_2$O$_7$, and
Sr$_4$Ru$_3$O$_{10}$ were obtained by varying the
SrCO$_3$:SrCl$_2$ ratio and temperature profiles. The temperature
profile was basically as follows: (1) Temperature increase to
1400-1500$^\circ$C in 7 hours and then constant temperature for 25
hours; (2) Cooling down to 1250-1350$^\circ$C at a rate of 2
deg/hour; (3)  Further cooling to room temperature in 1 hour.

The x-ray diffraction showed that the lattice parameters of all
three compounds corresponded to the ones known from the
literature. It was also established and further confirmed by Raman
polarization selection rules that the large surfaces of the
crystal platelets were parallel to the (001) plane and their edges
were along either  \{100\} or  \{110\} directions.

Raman spectra were collected using Jobin-Yvon HR640 spectrometer
 equipped with microscope (100$\times$ or 50$\times$ objective, focus spot size
1-3~$\mu$m), notch filters and liquid-nitrogen-cooled
charge-coupled device (CCD) detector. The He-Ne (632.8~nm) and
Ar$^+$ (514.5~nm and 488.0~nm) laser lines were used for
excitation. The lack of spurious signals from impurity phases was
verified by the reproducibility of the spectra and their strict
polarization. Given the crystallographic directions were known,
measurements could be done in several exact backward scattering
configurations: $z(xx)\bar{z}$, $z(yy)\bar{z}$, $z(xy)\bar{z}$,
$z(x'x')\bar{z}$, $z(x'y')\bar{z}$, $y(zz)\bar{y}$,
$y(zx)\bar{y}$, $y(xx)\bar{y}$, $y(x'x')\bar{y}$ . The first and
forth letters in these notations stay for the directions of
incident and scattered light, whereas their polarizations are
denoted by the second and third letters, respectively. As $x\equiv
[100]$ and $y\equiv [010]$ are indistinguishable, $x$ and $y$ are
interchangeable. The same is valid for $x'\equiv [110]$ and
$y'\equiv [1\bar{1}0]$. Further, the short notations $xx$, $zz$,
$x'x'$, $x'y'$ and $zx$ will also be used.

The lattice dynamical calculations were done using a shell model
described in detail in Ref.\cite{popov1}. This model gives an
adequate description of the vibrations in perovskitelike
structures because it accounts for their predominant ionicity. The
ionic interactions are represented by long-range Coulomb
potentials and short-range repulsive potentials of the Born-Mayer
form $ae^{-br}$ where $a$ and $b$ are constants and $r$ is the
interionic separation. The deformation of the electron charge
density of the ions is described in the dipole approximation
considering each atom as consisting of a point charged core and a
concentric spherical massless shell with charge $Y$. Each core and
its shell are coupled together with a force constant $k$ giving
rise to the free ionic polarizability $\alpha =  Y^2/k$. The model
parameters for the strontium, ruthenium, and oxygen ions and their
interaction potentials are taken from a previous study of simpler
compounds with perovskitelike structure\cite{popov1,iliev1}.

\section{Results and Discussion}

\subsection{Sr$_2$RuO$_4$}

The polarized Raman spectra of Sr$_2$RuO$_4$ at room temperature
as obtained with 488.0~nm excitation are shown in Fig.2. Three of
the four ($2A_{1g} +2E_g$) Raman allowed phonons are observed at
200~cm$^{-1}$ ($A_{1g}$, Sr vibrations along $z$), 247~cm$^{-1}$
($E_g$, apex oxygen vibrations in the $xy$ plane), and
545~cm$^{-1}$ ($A_{1g}$, apex oxygen vibrations along $z$), in
consistence with earlier reports of Udagawa et al.\cite{udagawa1}
and Sakita et al.\cite{sakita1}. The Raman phonon intensities
$I_\omega$ exhibit clear resonant behavior. For example in the
$zz$-polarized spectra the $I_{545}/I_{200}$ ratio is 19.3 for
632.8~nm (1.96~eV), 4.2 for 514.5~nm (2.41~eV), and 3.5 for
488.0~nm (2.54~eV) excitations, respectively. In the $xx$
polarized spectra the corresponding values are 2.1, 1.5, and 4.2.
For all three excitation energies used, the $y(xx)\bar{y}$ and
$z(xx)\bar{z}$ spectra were practically identical.

\begin{figure}
\includegraphics[width=3in]{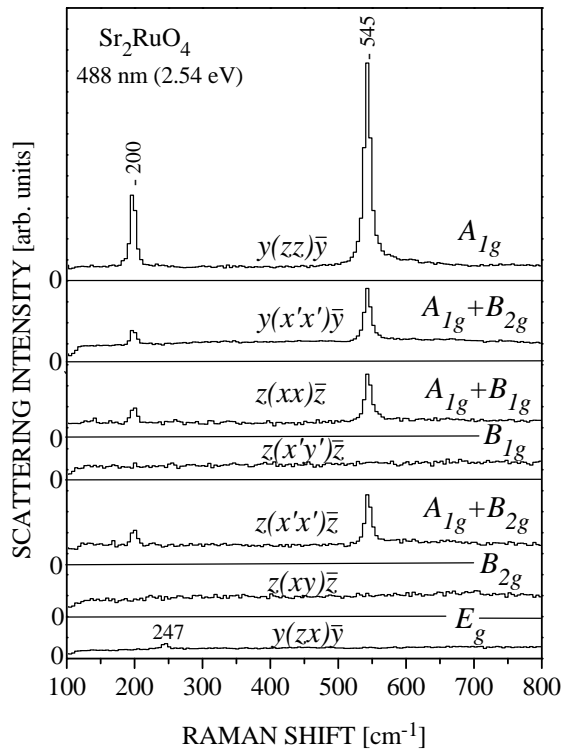}
\caption{Polarized Raman spectra of Sr$_2$RuO$_4$, obtained at
room temperature with 488~nm excitation. A continuum background is
present in all spectra.}
\end{figure}

Except for the phonon lines, an electronic background was present
in all spectra. The electronic Raman scattering for incident
polarization parallel to the $ab$ plane has previously been
reported by Yamanaka et al.\cite{yamanaka1}, who observed
structureless continuum with $A_{1g}$, $B_{1g}$ and $B_{2g}$
components of comparable intensity. Our measurements have shown,
however, that the $A_{1g}(xx)$ component of the continuum is much
weaker than those of $B_{1g}(x'y')$ and $B_{2g}(xy)$ symmetry.

\subsection{Sr$_3$Ru$_2$O$_7$}
The orthorhombic $Pban$ structure of Sr$_3$Ru$_2$O$_7$ can be
obtained from the idealized $I4/mmm$ structure (similar to
Sr$_2$RuO$_4$, see also Fig.1), by ordered counter-phase rotations
of RuO$_6$ octahedra around the $z$-axis. The $x$ and $y$ axes of
the $Pban$ structure are rotated by 45 degree with respect to
those of the tetragonal one and the $a$ and $b$-parameters are
larger by factor $\sqrt{2}$. While ten $\Gamma$-point phonon modes
($4A_{1g} + B_{1g} + 5E_g$) are Raman-allowed in the tetragonal
$I4/mmm$ structure of Sr$_3$Ru$_2$O$_7$, much more modes ($12A_g +
16B_{1g} + 22B_{2g} + 22B_{3g}$) are Raman-allowed in the $Pban$
structure due to the doubling of unit cell, absence of the cell
centering and appearance of new modes related to RuO$_6$
rotations. One should not expect, however, observation of such a
large number of Raman lines, as to each $A_{1g}$, $B_{1g}$ or
$E_g$ tetragonal mode one can juxtapose a pair of two $A_g$, two
$B_{1g}$ or $B_{2g} + B_{3g}$ orthorhombic modes, which involve
practically the same atomic vibrations and have very close
frequencies (see Table I for LDC of the $A_g$ and $B_{1g}$ modes).
Experimentally, these pairs will be observed as a single line.

\begin{figure}
\includegraphics[width=3in]{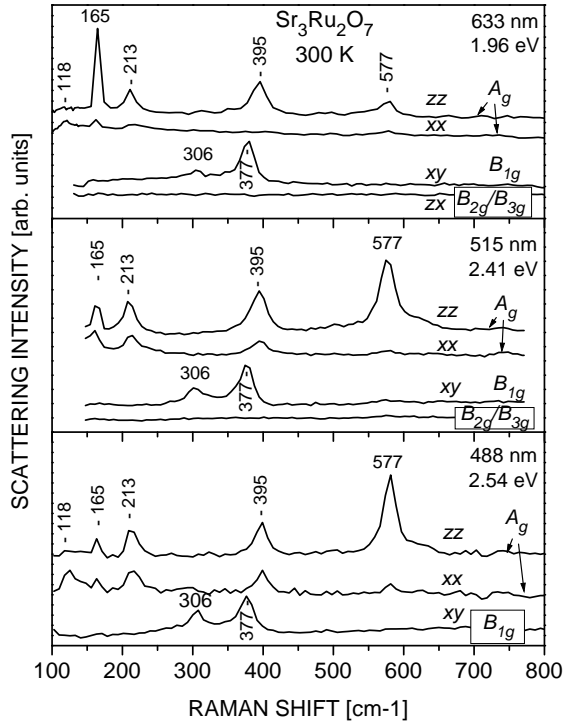}
\caption{Polarized Raman spectra of Sr$_3$Ru$_2$O$_7$, obtained at
room temperature with 633~nm, 515~nm, and 488~nm excitation. Some
curves are shifted vertically for clarity.}
\end{figure}

\begin{figure}
\includegraphics[width=2.5in]{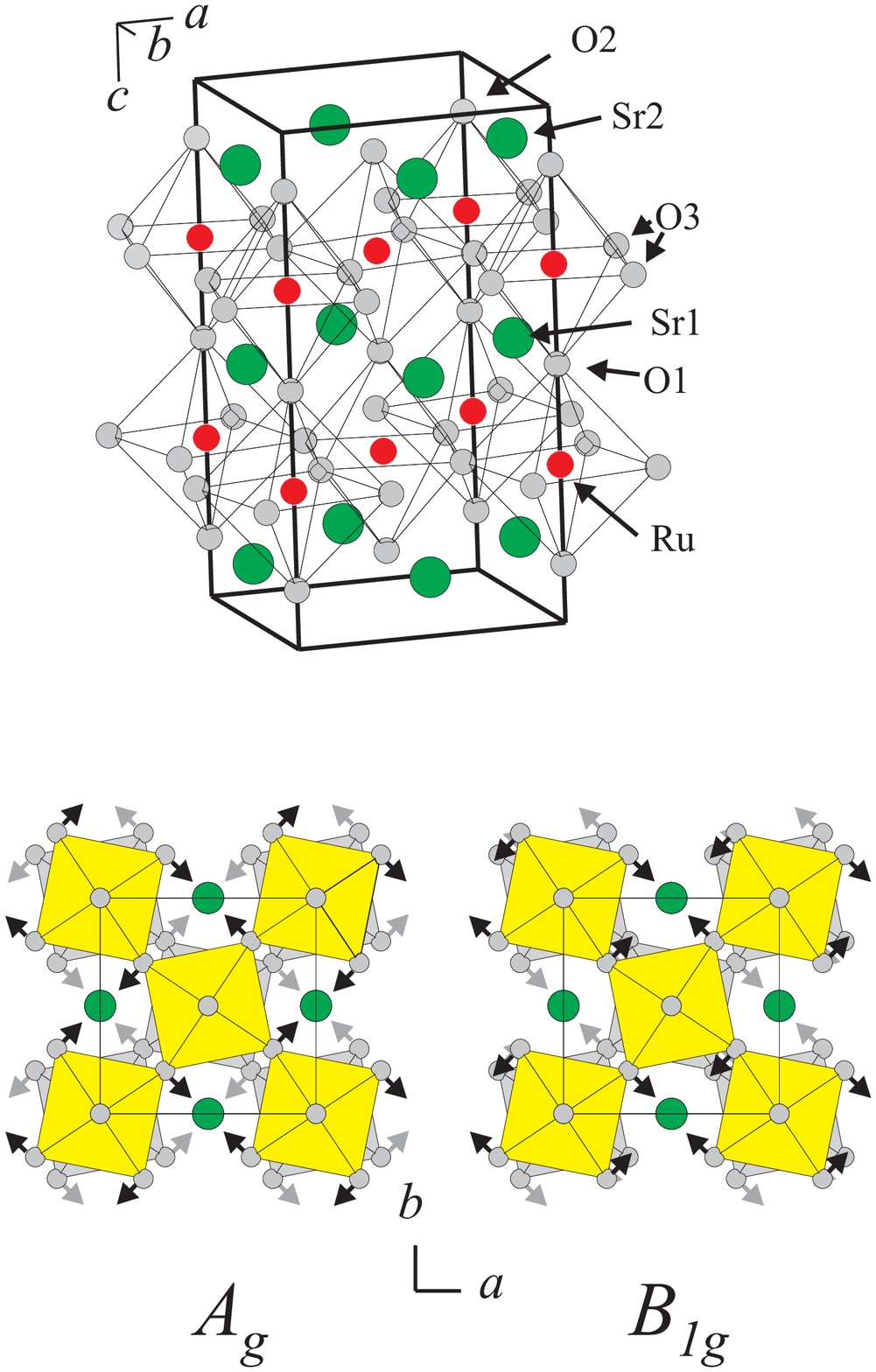}
\caption{Simplified distorted structure ($P4/nbm$) of
Sr$_3$Ru$_2$O$_7$ with elementary cell containing one double Ru-O
layer. The distortion-activated $A_g$ and $B_{1g}$ modes are also
shown.}
\end{figure}

\begin{figure}
\includegraphics[width=3.5in]{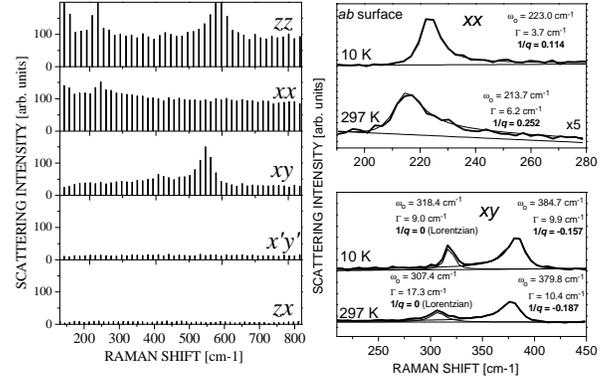}
\caption{Left panel: Continuum scattering from Sr$_3$Ru$_2$O$_7$
as obtained with 633~nm excitation for several exact scattering
configurations; Right panels: Fano fit of the 213/223~cm$^{-1}$
$A_g$~($xx$) and 377/384~cm$^{-1}$ $B_{1g}$ lines. The fit
parameters are also given.}
\end{figure}

Polarized Raman spectra of Sr$_3$Ru$_2$O$_7$ measured at room
temperature with 633~nm, 515~nm and 488~nm laser line excitation
are shown in Fig.~3. Five lines of $A_g$ symmetry are clearly
pronounced in the $xx$ and $zz$ spectra at 118, 165, 212, 395, and
577~cm$^{-1}$. Another two lines of $B_{1g}$ symmetry are seen in
the $xy$ spectra at 306 and 377~cm$^{-1}$, but no lines of
detectable intensity are observed in the $zx$($zy$) spectra where
are allowed the modes of $B_{2g}$($B_{3g}$) symmetry. The
intensity of the Raman lines changes with excitation energy due to
alteration of the resonance conditions. The relative intensities
of the $A_g$ line at 395~cm$^{-1}$ and the two $B_{1g}$ lines,
however, remain nearly the same. This suggests that the three
modes involve motions of same type of atoms, different from those
with main contribution to the $A_g$ mode at 577~cm$^{-1}$.  The
number of Raman lines in the spectra of Sr$_3$Ru$_2$O$_7$ is
noticeably lower than in the corresponding spectra of
Ca$_3$Ru$_2$O$_7$.\cite{liu1} This had to be expected as the
structure of Ca$_3$Ru$_2$O$_7$ is more strongly distorted and the
number of fully symmetrical modes allowed in $xx$, $yy$ and $zz$
configurations  is higher.

A simple approach to the  assignment of the Raman lines is based
on the reasonable assumption for weak interaction between the
double Ru-O slabs, the only important distortion being the
rotation of the RuO$_6$ octahedra. Within such assumption, instead
of the real $Pban$ structure, one can consider a simplified
structure with elementary cell containing only one double layer
(Fig.4). This structure is tetragonal ($P4/nbm$, No.125, Z=2) with
the same $a$ and $b$ parameters as the real one, but twice shorter
$c$ parameter. The normal mode analysis gives $5A_{1g} + B_{1g} +
4B_{2g} + 11E_g$ Raman modes. Therefore one expects in the $xx$
and $zz$ spectra five Raman lines of $A_{g}$ symmetry ($A_{1g}$ in
tetragonal notations), which is exactly the case. In the $xy$
spectra one expects observation of four $B_{1g}$ modes ($B_{2g}$
in tetragonal notations), two of them involving mainly oxygen
vibrations.

The Raman mode frequencies in ionic materials, such as transition
metal oxides, are determined by the mass, charge, and bond lengths
of participating atoms as well by the type of atomic motions
(stretching, bending or rotational). Based on comparison to other
perovskitelike oxides, the modes involving mainly vibration of
heavier Ru and Sr or rotational vibrations of oxygens are expected
in the frequency range below 250~cm$^{-1}$, the bending oxygen
modes - between 200 and 500~cm$^{-1}$ and stretching oxygen modes
- above 500~cm$^{-1}$. Our LDC for the $Pban$ structure (Table I)
predict that the $A_g$ modes below 250~cm$^{-1}$ are strongly
mixed, each involving vibrations along $z$ of Ru and Sr2 as well
as RuO$_6$ rotations around $z$. In the $I4/mmm$ structure,
however, the rotational motions (see Fig.4), are not Raman active.
Therefore, the 165~cm$^{-1}$ and 213~cm$^{-1}$ lines, which are
close to LDC frequencies predicted for both $Pban$ and $I4/mmm$
structures, can tentatively be assigned to mixed vibrations of Ru
and Sr along $z$. The line at 118~cm$^{-1}$ is close to the
LDC($Pban$) frequency of 117~cm$^{-1}$, which has no partner in
the LDC($I4/mmm$) data, is assigned to mainly RuO$_6$ rotations.
The "soft"-mode temperature behavior of the latter line is also
typical for a rotational mode. As to the two high-frequency $A_g$
modes, with great certainty, confirmed by LDC, they correspond to
out-of-plane in-phase vibrations of O3 (395~cm$^{-1}$) and
stretching vibrations of O2 (577~cm$^{-1}$). The latter frequency
is higher that that of the corresponding apex oxygen vibrations in
Sr$_2$RuO$_4$ (545~cm$^{-1}$), which can be explained accounting
the bond-length changes. Indeed, it is plausible to assume that
the force constants $k_{\rm O-M}$ between O and the ion M follows
the simple $Z_{\rm O}Z_{\rm M}/r_{\rm O-M}^3$ relation, where
$Z_{\rm O}$ and $Z_{\rm M}$ are the charges of the oxygen and
cation, respectively, and $r_{\rm O-M}$ is the oxygen-cation bond
length. This relation is valid for harmonic ionic crystals and
shown to apply for perovskitelike transition metal
oxides.\cite{kakihana1,hadjiev1}. Taking into account that
$\omega^2 = k/m$ and restricting interactions to only
nearest-neighbors (Ru and Sr), one obtains for the O2 stretching
vibrations
\begin{equation}
\omega^2 \propto  Z_{\rm Ru}/r_{\rm O-Ru}^3 + Z_{\rm Sr}/r_{\rm
O-Sr}^3
\end{equation}
where $Z_{\rm Ru}=4$, $Z_{\rm Sr}=2$, and the values of $r_{\rm
O-Ru}$ and $r_{\rm O-Sr}$ are respectively 2.016 \AA \ and 2.459
\AA \ for Sr$_3$Ru$_2$O$_7$\cite{huang1} and 2.059 \AA \ and 2.440
\AA \ for Sr$_2$RuO$_4$\cite{neumeier1}. Using these values one
obtains $\omega_{\rm Sr_3Ru_2O_7}/\omega_{\rm Sr_2RuO_4} = 1.045$,
which is close to the experimental  ratio of 1.059.

 There is little doubt too, that the two
experimentally observed lines in the $xy$ spectra correspond to
the two oxygen $B_{2g}$ modes of the distorted $P4/nbm$ structure.
The main atomic motions are out-of-phase vibrations along $z$ of
O3 (306~cm$^{-1}$) and "scissors-like" bendings of O3 parallel to
the $xy$ plane (377~cm$^{-1}$). The shape of the latter mode is
also shown in Fig.4.

Along with the phonon lines, a structureless background of
$A_g$($zz$ and $xx$), and $B_{1g}$ ($xy$) symmetry, but not of
$B_{2g}/B_{3g}$ ($zx$ or $zy$) symmetry, is also present in the
spectra. This is illustrated in Fig.5 for room temperature spectra
taken with 633~nm excitation. The observation of relatively strong
$zz$-polarized continuum was somewhat unexpected as the ARPES
measurements\cite{puchkov1} and band structure
calculations\cite{singh1} provide evidence for
quasi-two-dimensional Fermi-surface sheets, in consistence with
reports for large anisotropy of the electrical resistivity
($\rho_z/\rho_{xy} \approx 40$ at 300~K)\cite{ikeda1}. The
continuum-phonons interference is clearly pronounced only for
incident radiation parallel to the $xy$ plane through Fano shape
of the 213~cm$^{-1}$ $A_g$ line in the $xx$ spectra and the
377~cm$^{-1}$ $B_{1g}$ line in the $xy$ spectra (Fig.5). For a
phonon coupled to electronic background, the Fano profile
$I(\omega) = I_0 (\epsilon + q)^2/(1+\epsilon^2)$ is generally
used to describe the line shape, where $\epsilon = (\omega -
\omega_0)/\Gamma$, $\omega_0$ is the "bare" phonon frequency,
$\Gamma$ is the linewidth, and $q$ is the asymmetry parameter,
$1/q \propto V$ reflects the electron-phonon interaction $V$.
While for an uncoupled phonon $1/q = 0$, the increase of
electron-phonon interaction increases $|1/q|$. The values of
$\omega_0$, $\Gamma$, and $1/q$ as obtained from the Fano fit of
the experimental line profiles are listed in Fig.5.

\begin{figure}
\includegraphics[width=3in]{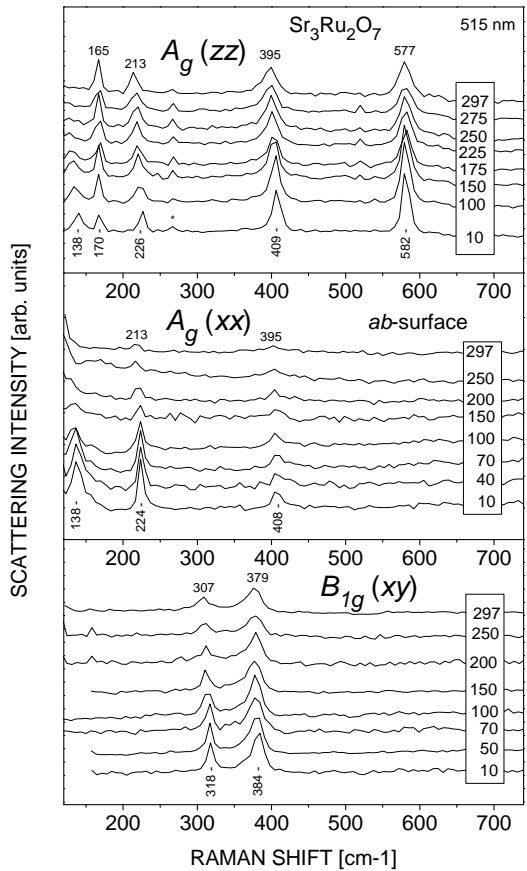}
\caption{Temperature-dependent polarized Raman spectra of
Sr$_3$Ru$_2$O$_7$ between 10 and 300 K obtained with 515~nm
excitation.}
\end{figure}

\begin{figure}
\includegraphics[width=2.5in]{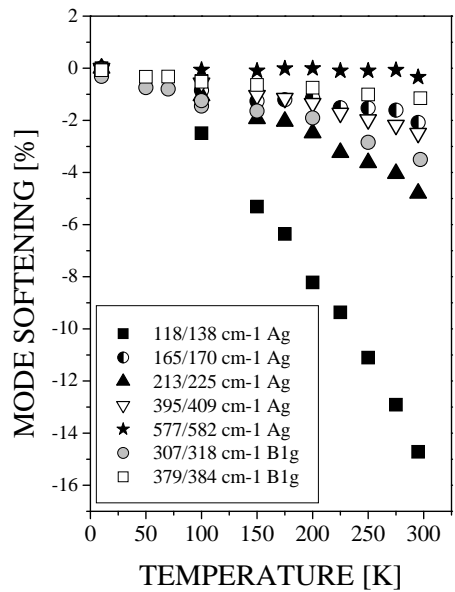}
\caption{Variations with T of the relative changes
$[\omega(T)-\omega(10)]/\omega(10)$ of the $A_g$ mode frequencies
in Sr$_3$Ru$_2$O$_7$.}
\end{figure}

Fig.6 shows the temperature-dependent $zz$, $xx$ and $xy$ Raman
spectra of Sr$_3$Ru$_2$O$_7$ between 10 K and 300 K. The
variations of phonon frequencies and line widths with temperature
exhibit no anomalies. The relative changes of frequency for all
observable modes are summarized in Fig.7. The low frequency $A_g$
mode shows "soft"-mode behavior, its frequency decreasing by
$\approx 15\%$ between 10 and 300~K, while the frequency of the
highest mode remains practically unchanged. The continuum
scattering in the $zz$, $xx$ and $xy$ spectra only slightly
decreases with lowering temperature,  in contrast to the case of
Ca$_3$Ru$_2$O$_7$, where a rapid suppression of the electronic
background of $B_{1g}$ symmetry has been observed below the
metal-insulator transition at $T_{MI}\approx 48$~K.\cite{liu1} The
interaction between the continuum and phonons also decreases with
lowering temperature, as evidenced from Fig.5, where are compared
the $1/q$ values at 10 and 300~K.

\begin{table}[tbp]
\caption{Calculated (for the $Pban$ and $I4/mmm$ structures) and
experimentally observed $A_g/A_{1g}$ and B$_{1g}$ phonon
frequencies in Sr$_3$Ru$_2$O$_7$.}
\begin{tabular}{|cccc|cc|c|}
\hline

Mode& LDC&   Exp&Atomic &Mode &LDC& Mode \\
$Pban$&$Pban$& 300K/10K& motions & \multicolumn{2}{c}{ $I4/mmm$}& $P4/nbm$  \\
\hline
$A_g(1)$&117&118/138 &RuO$_6$ rot& & & $A_{1g}$ \\
 $A_g(2)$&179&165/170&Ru($z$), Sr2($z$) & $A_{1g}$ & 180& $A_{1g}$ \\
 $A_g(3)$&210&213/225 &Sr2($z$), Ru($z$)  &$A_{1g}$ &214 &$A_{1g}$  \\
  $A_g(4)$&223&213/225& & & & \\
 $A_g(5)$&250& &             & & & \\
 $A_g(6)$&254& &             & & & \\
 $A_g(7)$&402&395/409 & O3($z$) &$A_{1g}$ & 512& $A_{1g}$\\
 $A_g(8)$&402&395/409 & & & &\\
 $A_g(9)$&520 & & & & & $$\\
 $A_g(10)$&522 & & & & & \\
$A_g(11)$&580 &577/581& O2($z$) &$A_{1g}$ &589 & $A_{1g}$\\
$A_g(12)$&580 &577/581 &  &  & & \\
\hline
$B_{1g}(1)$&108 &     &  & &  & \\
$B_{1g}(2)$&109 &     &  & &  & \\
$B_{1g}(3)$&162 &     &  & &  & $B_{2g}$\\
$B_{1g}(4)$&162 &     &  & &  & \\
$B_{1g}(5)$&181 &     &  & & & $B_{2g}$\\
$B_{1g}(6)$&181 &     &  & &  & \\
$B_{1g}(7)$&337 &306/316 & O3($z,-z$) &$B_{1g}$ &310 & $B_{2g}$ \\
$B_{1g}(8)$&337 &306/316 &   & & & \\
$B_{1g}(9)$&378 &377/380 & O3($xy$) & & & $B_{2g}$\\
$B_{1g}(10)$&378 &377/380 &  & & & \\
$B_{1g}(11)$&466 &     &  & & & \\
$B_{1g}(12)$&466 &     &  & & & \\
$B_{1g}(13)$&521 &     &  & & & \\
$B_{1g}(14)$&521 &     &  & & & \\
$B_{1g}(15)$&726 &     &  & & & \\
$B_{1g}(16)$&726 &     &  & & & \\
\hline
\end{tabular}
\end{table}

\subsection{Sr$_4$Ru$_3$O$_{10}$}

The structure of Sr$_4$Ru$_3$O$_{10}$ has been refined as
$Pbam$.\cite{crawford1} From symmetry considerations in total 96
($20A_g + 20B_{1g} + 28B_{2g} + 28B_{3g}$) modes are Raman
allowed. Like in the case of Sr$_3$Ru$_2$O$_7$, however, most
modes are practically degenerated in pairs and lower number of
Raman lines is expected in the spectra. Fig.8 shows the $zz$,
$xx$, $xy$ and $zx$ spectra obtained at room temperature with
633~nm, 515~nm, and 488~nm excitation. The temperature variations
of the spectra between 10 and 300~K are given in Fig.9. Except for
line shifts and appearance of additional weak lines, the spectral
profiles and their dependence on  scattering configuration and
excitation wavelength resembles that of Sr$_3$Ru$_2$O$_7$. Lines
of $A_g$ symmetry are observed at room temperature 129, 163,
195-203, 360-365, 533, 582-585, 619 and 745~cm$^{-1}$. The broad
structure near 200~cm$^{-1}$ in the $xx$ spectra appears to be a
superposition of two lines. This becomes evident at low
temperatures where these two line are well separated. Like in the
case of Sr$_3$Ru$_2$O$_7$, only two lines of $B_{1g}$ symmetry are
pronounced and the intensity of the $B_{2g}$ and $B_{3g}$ lines is
below the detection limit.

\begin{figure}
\includegraphics[width=3in]{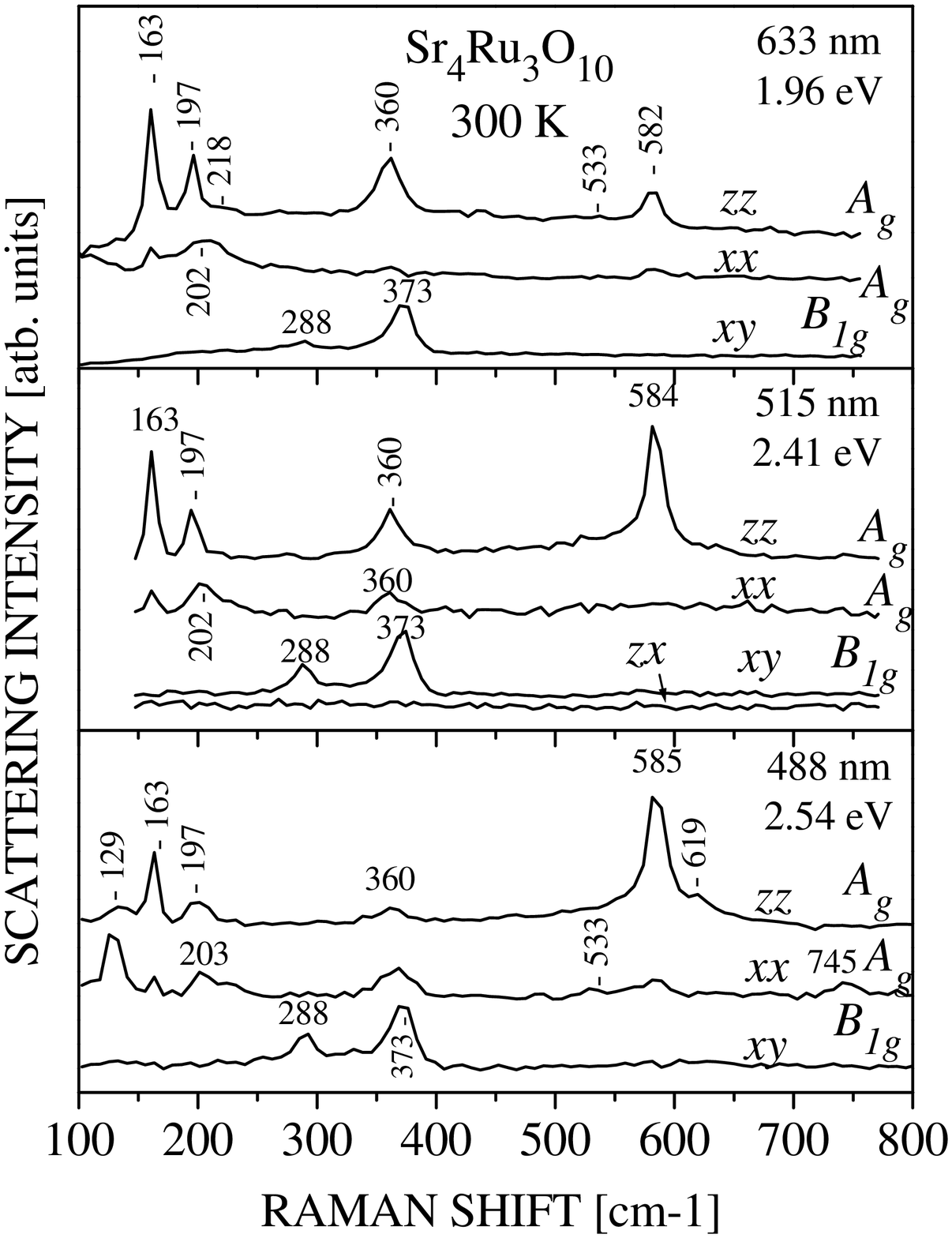}
\caption{Polarized Raman spectra of Sr$_4$Ru$_3$O$_{10}$, obtained
at room temperature with 633~nm, 515~nm, and 488~nm excitation.
Some curves are shifted vertically for clarity.}
\end{figure}

\begin{figure}
\includegraphics[width=3in]{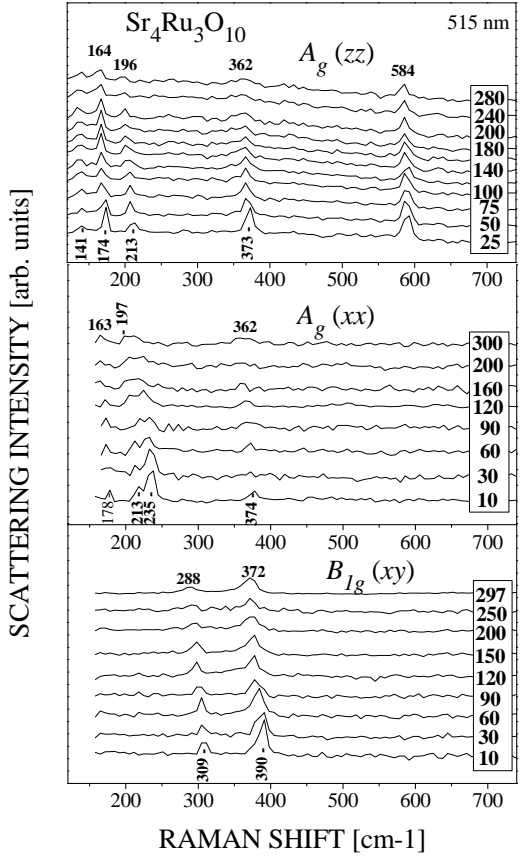}
\caption{Variation with T of the $zz$, $xx$ and $xy$ spectra of
Sr$_4$Ru$_3$O$_{10}$.}
\end{figure}

\begin{figure}
\includegraphics[width=2in]{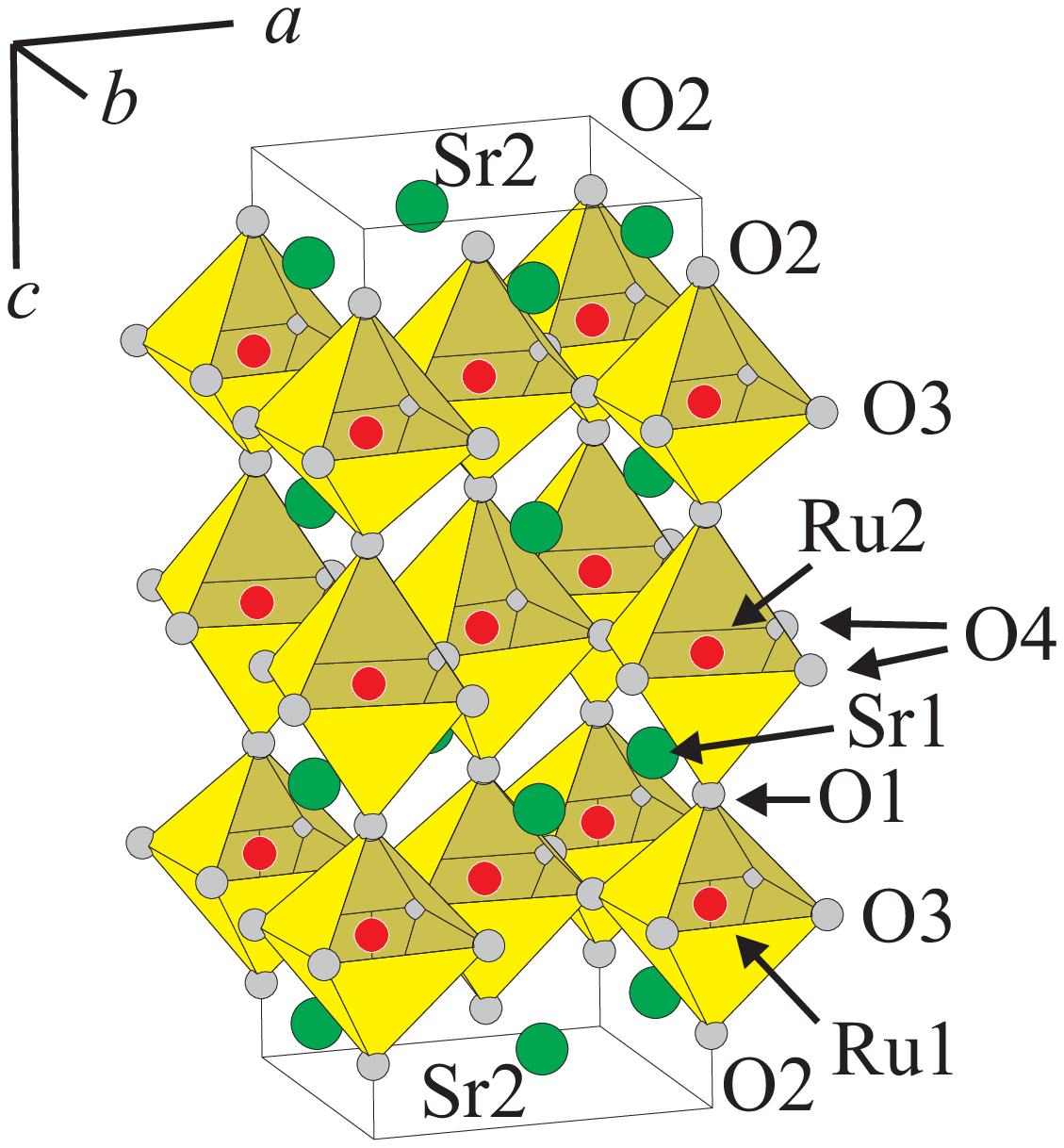}
\caption{Simplified distorted structure of Sr$_4$Ru$_3$O$_{10}$:
$P4/mbm$, No.127, Z=2.}
\end{figure}

\begin{figure}
\includegraphics[width=3.4in]{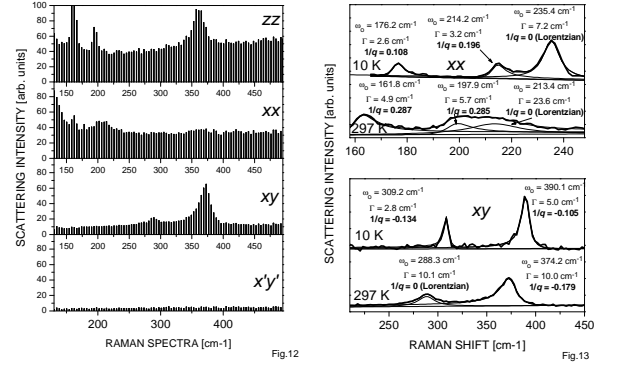}
\caption{Left panel: Continuum scattering in the spectra of
Sr$_4$Ru$_3$O$_{10}$ obtained with 633~nm excitation; Right
panels: Fano fits $xx$($A_g$) and $xy$($B_{1g}$) spectra of
Sr$_4$Ru$_3$O$_{10}$ obtained with 515 nm excitation at 10~K and
300~K. The fit parameters are also given.}
\end{figure}

To assign the Raman lines to particular phonon modes let us again
consider the Raman modes in a simplified distorted tetragonal
structure and compare the experimental frequencies to the
prediction of lattice dynamical calculations for the real $Pbam$
structure of Sr$_4$Ru$_3$O$_{10}$. The simplified structure
$P4/mbm$ (No.127, Z=2), shown in Fig.10, has only one triple layer
in the unit cell. It is characterized by the same equal $a$ and
$b$ parameters as the real structure, but twice shorter $c$
parameter. From symmetry considerations $8A_{1g} + 2B_{1g} +
5B_{2g} + 14E_g$ modes are Raman active. The $A_{1g}$ modes,
allowed in $xx$ and $zz$ spectra, correspond to $A_g$ modes in
orthorhombic $Pbam$, whereas the tetragonal $B_{2g}$ modes,
allowed in the $xy$ spectra, correspond to the orthorhombic
$B_{1g}$ modes. Compared to the simplified structure of
Sr$_3$Ru$_2$O$_7$, there are three new $A_{1g}(A_g)$ modes
corresponding to: (i) Rotations of middle RuO$_6$ octahedra; (ii)
Vibrations along $z$ of internal apex oxygen atoms (O1); (iii)
Vibrations along $z$ of internal Sr1 atoms. The LDC predict close
frequencies for the rotational and Sr1 modes and our assignment of
the new $xx$ polarized $A_g$ line, seen at low temperature at
235~cm$^{-1}$ and also having a "soft"-mode temperature behavior,
to rotational vibrations of middle RuO$_6$ octahedra is only
tentative. There is little doubt that the new $A_g$ lines at
745~cm$^{-1}$, seen in the $xx$ spectra with 488~nm excitation,
correspond to the vibrations along $z$ of the inner O1 atoms.
Using Eq.(1) with the experimental values of $r_{O1-Ru2} =
1.991$~\AA \ , $r_{O1-Ru1} = 2.014$~\AA \  and $r_{O2-Ru1} =
2.077$~\AA \ , $r_{O2-Sr2'} = 2.380$~\AA \ one obtains frequency
of O1 vibrations along $z$ should be by factor 1.35 higher than
that of the outer apex oxygens (O2). This is in good agreement
with the experimentally observed ratio of 745:584 = 1.28. A
comparison of phonon frequencies predicted by LDC with
experimental data is given in Table II.

\begin{table*}[tbp]
\caption{Calculated and experimentally observed $A_{1g}$ and
B$_{1g}$ phonon frequencies in Sr$_4$Ru$_3$O$_{10}$. The
corresponding modes in the simplified $P4/mbm$ structure are also
given.}
\begin{tabular}{|cccc|c|cccc|c|}
\hline

Mode& LDC&   Exp&Atomic &Mode &Mode &LDC& Exp     & Atomic &Mode \\
$Pbam$&$Pbam$& 300K/10K& motions  &$P4/mbm$ &$Pbam$&$Pbam$& 300K/10K& motions&$P4/mbm$  \\
\hline
$A_g^{(1)}$ & 90 & 129/140   & outer RuO$_6$ rot in $xy$ & $A_{1g}$&$B_{1g}^{(1)}$&147&    & &    \\
$A_g^{(2)}$ &150 & 163/178    & Ru1($z$)Sr2($z$)&$A_{1g}$ & $B_{1g}^{(2)}$&151&   &   &  \\
$A_g^{(3)}$ &173 &    &  &       & $B_{1g}^{(3)}$&175  &      & &$B_{2g}$ \\
$A_g^{(4)}$ &190 & 198/214    & Sr2($z$)Ru1($z)$ & $A_{1g}$& $B_{1g}^{(4)}$&176 &      &  &\\
$A_g^{(5)}$ &198 &     &  &         & $B_{1g}^{(5)}$& 185    &       & &$B_{2g}$  \\
$A_g^{(6)}$ &210 &            &  &  & $B_{1g}^{(6)}$& 188     & &  & \\
$A_g^{(7)}$ &231 &213/235     & middle RuO$_6$ rot in $xy$ &$A_{1g}$ & $B_{1g}^{(7)}$&327       & 288/307   &O3($z,-z$)  &$B_{2g}$  \\
$A_g^{(8)}$ &234 &     &  &         & $B_{1g}^{(8)}$     & 327  &   & &  \\
$A_g^{(9)}$ &286 &          &              &       & $B_{1g}^{(9)}$     & 354     &373/388    & O3($xy$)  &$B_{2g}$\\
$A_g^{(10)}$ &288&          &            &         & $B_{1g}^{(10)}$     &356     &    &  & \\
$A_g^{(11)}$ &309&          &   & & $B_{1g}^{(11)}$     &464      &       &   &$B_{2g}$\\
$A_g^{(12)}$ &313&          &  &         & $B_{1g}^{(12)}$     &464      &       &   & \\
$A_g^{(13)}$ &358&361/372   & O3($z$)  &$A_{1g}$   & $B_{1g}^{(13)}$     &476      &       &   & \\
$A_g^{(14)}$ &364&   &  &              & $B_{1g}^{(14)}$     &476      &       &   & \\
$A_g^{(15)}$ &512&533/      &O3($xy$) & $A_{1g}$& $B_{1g}^{(15)}$     &482      &       &   &  \\
$A_g^{(16)}$ &514&      &  & & $B_{1g}^{(16)}$     &482      &       &   &  \\
$A_g^{(17)}$ &552& 584/588  & O2($z$)  & $A_{1g}$                & $B_{1g}^{(17)}$     &708      &       &  &  \\
$A_g^{(18)}$ &562&   &      &                & $B_{1g}^{(18)}$     &708      &       &    & \\
$A_g^{(19)}$ &678& 745/     &O1($z$)            &$A_{1g}$             &      &      &       &   & \\
$A_g^{(20)}$ &678&     & & &             &      &      &       &    \\
\hline
\end{tabular}
\end{table*}

The left panel of Fig.11 illustrates the presence in the $zz$,
$xx$ and $xy$ spectra of Sr$_4$Ru$_3$O$_{10}$ of scattering
continuum. Like in the case of Sr$_3$Ru$_2$O$_7$, some lines
exhibit clear Fano shape for light polarization parallel to the
$ab$ plane. This is shown in more detail in the right panels of
Fig.11.

The ferromagnetic ordering at T=105~K has a moderate, but clearly
pronounced effect on phonon parameters and electron-phonon
interaction. The temperature dependencies of the position
(Fig.12a) and width (Fig.12b) of the Fano shaped $B_{1g}$ line,
corresponding to O3 vibrations in the $xy$ plane,  change their
slope near T$_C$. The changes of $1/q(T) \propto V(T)$ (Fig.12c)
are even more pronounced, providing evidence that the
electron-phonon coupling decreases in the ferromagnetic phase.
Weaker changes near T$_C$ of the parameters of the second $B_{1g}$
line, corresponding to O3 vibrations in $z$ direction, are also
observed (Figs.12d and 12e).

The magnetic ordering may affect phonon frequency through
different mechanisms: exchange striction,
\cite{baltensperger1,udagawa1} dependence of the spin energy on
ion displacements\cite{baltensperger1,udagawa2,chen1}, variations
of the density of itinerant carriers near T$_M$.\cite{iliev1}
Although consistent with observation of anomaly in $1/q$, the
latter mechanism, proposed to explain the anomalous hardening near
T$_C$ of several Raman modes in ferromagnetic SrRuO$_3$, is
irrespective to the bond lengths and band angles being modulated
and seems to be of less importance in Sr$_4$Ru$_3$O$_{10}$. The
observation of detectable anomaly near T$_C$ only for modes, which
modulate the Ru1-O3-Ru1 angle, rather favors direct interaction of
these phonons with the magnetization.\cite{baltensperger1, chen1}.

\begin{figure}
\includegraphics[width=2in]{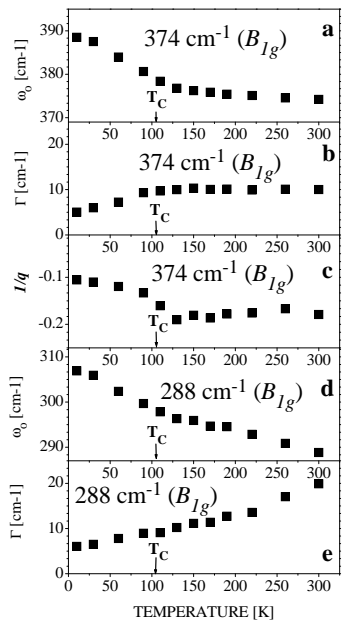}
\caption{Variations with T of (a): $\omega_0$ ,(b): $\Gamma$ , and
(c): $1/q \sim V$ for the Fano shaped $B_{1g}$ line at
374/389~cm$^{-1}$. No anomaly near T$_C$ is observed, however, for
the position (d) and the width (e) of the  Lorentzian $B_{1g}$
line at 288/308~cm$^{-1}$.}
\end{figure}

\section{Conclusions}

We studied in detail the polarized Raman spectra of Sr$_2$RuO$_4$,
Sr$_3$Ru$_2$O$_7$ and Sr$_4$Ru$_3$O$_{10}$ with particular
attention to the two latter compounds as their phonon spectra have
not been reported so far. All observed Raman lines are of either
$A_g$ or $B_{1g}$ symmetry. They have been assigned to definite
atomic vibrations by: (1) considering the Raman active modes in
simplified tetragonal $P4/nbm$ and $P4/mbm$ structures, which
contain only one double or triple RuO$_6$ layers per unit cell and
account for the rotational distortions; (2) comparison to the
predictions of lattice dynamical calculations for the real
orthorhombic double layer $Pbna$ and triple layer $Pbam$
structures. Except for the discrete phonon lines, a continuum
background, presumably of electronic origin, is present in the
$zz$, $xx$ and $xy$, but not in the $x'y'$ and $zx$ spectra. The
interaction of $xx$ and $xy$ continuum with the modes involving
atomic motions in the $ab$ plane results in Fano shape of
corresponding Raman lines. While no anomaly in phonon parameters
of Sr$_3$Ru$_2$O$_7$ is seen between 10 and 300~K, where no
magnetic transition occur, an anomaly is observed near
ferromagnetic transition at 105~K in Sr$_4$Ru$_3$O$_{10}$.

\acknowledgments This work is supported in part by the state of
Texas through the Texas Center for Superconductivity and Advanced
Materials, by NSF grant no. DMR-9804325, the T.L.L. Temple
Foundation, the J. J. and R. Moores Endowment, and at LBNL by the
Director, Office of Energy Research, Office of Basic Energy
Sciences, Division of Materials Sciences of the US Department of
Energy under contract no. DE-AC03-76SF00098. J.B. acknowledges
financial support from the Swedish Superconductivity Consortium.

\end{document}